Spin Transfer Dynamics in Spin Valves with Out-of-plane Magnetized CoNi Free Layers

William H. Rippard, Alina M. Deac, Matthew R. Pufall, Justin M. Shaw, Mark W. Keller, Stephen E. Russek

*National Institute of Standards and Technology, Boulder, CO 80305, USA*

Gerrit E. W. Bauer

*Kavli Institute of NanoScience, Delft University of Technology, 2628 CJ Delft, The Netherlands*

Claudio Serpico

*Dipartimento di Ingegneria Elettrica, Università di Napoli "Federico II," 80125 Napoli, Italy*

Abstract:

We have measured spin transfer-induced dynamics in magnetic nanocontact devices having a perpendicularly magnetized Co/Ni free layer and an in-plane magnetized CoFe fixed layer. The frequencies and powers of the excitations agree well with the predictions of the single-domain model and indicate that the excited dynamics correspond to precessional orbits with angles ranging from zero to 90º as the applied current is increased at a fixed field. From measurements of the onset current as a function of applied field strength we estimate the magnitude of the spin torque asymmetry parameter $\Lambda \approx 1.5$. By combining these with spin torque ferromagnetic resonance measurements, we also estimate the spin wave radiation loss in these devices.

The spin transfer effect has been shown to give rise to coherent magnetization dynamics in a variety of magnetic devices and materials [1]. Most of the initial work focused on the dynamics excited for in-plane magnetized films, although out-of-plane magnetization dynamics have been induced through the use of applied magnetic fields [2,3]. More recently, it has been demonstrated that out-of-plane precession of in-plane magnetized films can be generated in device structures incorporating perpendicularly magnetized materials as the polarizing layer as well. However, such structures require an additional read-out layer in order to detect the oscillations through the giant magnetoresistance (GMR) effect [4,5] . This additional layer has the drawback of complicating the spin-dependent transport in the devices. Here, we show that by utilizing an *in-plane polarizing* layer and a *perpendicularly magnetized* Co/Ni free layer [6], coherent out-of-plane dynamics can be generated, having microwave output power on the order of 1 nW, without the need for a third read-out layer. While this orientation of the free and fixed layer magnetizations can be induced through applying out-of-plane magnetic fields to in-plane magnetized films, the ensuing spin transfer induced dynamics of a perpendicularly magnetized free layer are significantly different. In particular, the precession frequencies decrease with increasing current and are amenable to more detailed comparison to analytic theory.

Structures utilizing perpendicularly magnetized free layers have the additional advantageous feature that dynamics can be induced using applied fields that are relatively low compared to the saturation magnetization. This allows investigation of the dynamics of the free layer in changing applied fields while keeping the orientation of the fixed layer essentially constant, which significantly simplifies comparison to analytic theory. From these comparisons, we find that frequencies of the excitations and the associated device output power are both in accordance with the angle of precession (*i.e.* precessional amplitude) increasing from near zero to 90º with increasing current at a fixed applied field. Furthermore, we can estimate the magnitude of amplitude variations of the oscillations from the measured line widths at a fixed external field and current. From measurements of the onset current for dynamics vs. applied field, we estimate values for both the energy loss associated with spin wave radiation away from the active device area and the asymmetry parameter in the angular dependence of the spin torque.

The spin transfer nanocontact oscillator (STNO) devices studied here [3] consist of a nominal 70 nm diameter electrical contact to a 8 μm x 24 μm spin-valve mesa composed of substrate|Ta(3)|Cu(15)|Co$_{90}$Fe$_{10}$(20) |Cu(4.5)|[Co(0.2)|Ni(0.4)]$^{x5}$|Co(0.3)|Cu(3)|Ta(3) where the thicknesses are given in nanometers. Magnetometry measurements show that this Co|Ni multilayer has an out-of-plane magnetization with $\mu_0(H_k - M_s) \approx 0.068$ T, where $H_k$ is the intrinsic anisotropy perpendicular to the film plane and $M_s$ is the saturation magnetization, which has also been confirmed through GMR measurements and the ferromagnetic resonance data discussed below. The CoFe layer acts as the fixed layer, due to its larger value of $M_s$ and thickness, and the Co|Ni multilayer corresponds to the free layer in which dynamics are induced. The devices are DC current biased so that changes in the device resistance associated with precessional motion of the free layer magnetization result in a time varying voltage across the device through the GMR effect, which is measured with a spectrum analyzer. All measurements were performed at room temperature. The results presented here are from a single device but have been observed in tens of samples having similar compositions and values of $H_k - M_s$.

In Fig. 1a we show typical spectral output from a device as a function of DC bias for an external field $\mu_0 H = 0.25$ T applied perpendicular to the film plane. This field orientation more strongly saturates the free layer magnetization in this direction while pulling the fixed CoFe layer magnetization roughly 7º out of the film plane (assuming $\mu_0 M_s$ of CoFe is 1.8 T). As seen from the figure, the onset current $I_c$ for this device is 3.8 mA (positive currents correspond to electrons flowing from the free layer (Co/Ni) to the fixed layer (CoFe)). Near onset the spectral output consists of a single peak at a frequency $f = 9.3$ GHz that decreases (red-shifts) with increasing current. This red-shifting is distinctly different from the situation when an out-of-plane applied magnetic field is used to induce a similar magnetic configuration for films without significant perpendicular anisotropy, in which case the spin torque induces dynamics that blue-shift with increasing current [7,8]. As the current through the device $I_{dc}$ increases, a second harmonic signal appears ($I_{dc} \approx 8$ mA), indicating that the excited mode qualitatively changes as the highest current values are reached. This behavior is discussed in more detail below. Figures 1 (b,c) show the oscillation frequency and line width $\Delta f$ (corresponding to the full-width-at-half-maximum (FWHM), as determined by Lorentzian fits to the spectral peaks in part (a). The line width at onset is comparatively large, decreases with increasing current to a minimum of 6 MHz near $I_{dc}$ = 6 mA, and then increases at higher current values. We attribute the initial decrease in the line width is a general characteristic of STNO devices and results from both the spin torque and damping being small close to onset, leaving the oscillator susceptible to thermal fluctuations. The increase in the line width at the highest currents roughly correlates with the appearance of the second harmonic signal, and again indicates a qualitative change in the precessional motion in this regime.

In the following paragraphs we compare our experimental results with those from single-domain modeling. While the single-domain model only approximates the nanocontact devices, in which the active magnetic area is exchange coupled to the surrounding film, this comparison offers significant insight to the device behavior by capturing the general characteristics of the dynamics, thereby allowing for analytic modeling. In the single-domain modeling, we assume an out-of-plane magnetized free layer having $\mu_0 M_s$ = 1 T, a perpendicular anisotropy $H_k$ =1.1 $M_s$, and a fixed layer that is strictly magnetized in the plane of the film. This approximates the experiments in which $H_k$ = 1.07 $M_s$ and the applied fields cant $\boldsymbol{M_{fix}}$ out of the film plane up to about 11°. Micromagnetic modeling of this basic device configuration has been performed [9] and shows qualitative agreement with the measured red-shifting dynamics. However, these results are insufficiently detailed to allow for quantitative comparison with the present measurements.

Because the fixed layer is taken to be in the film plane, within the single-domain model an asymmetric angle dependence of the spin torque is essential for dynamics to be induced. The torque asymmetry $g(\psi)$ is taken to be $P\Lambda^2 \,[(\Lambda^2+1)+(\Lambda^2-1)\cos(\psi)]^{-1}$, where $P$ is the spin-current polarization, $\psi$ is the angle between $\mathbf{M}_{free}$ and $\mathbf{M}_{fixed}$, and $\Lambda$ is the asymmetry parameter.[10] In the case of a symmetric $g(\psi)$, ($\Lambda = 1$), the energy transferred by the spin torque to the free layer during one half of the precessional cycle is equivalently extracted during the other half of the cycle, and steady state dynamics are not sustainable. In the presence of an asymmetric $g(\psi)$ and $\Lambda > 1$, the energy transferred to the free layer during one half of its precessional cycle is not equally extracted during the other half and there is a net transfer of energy by the spin torque. For one sign of current a positive energy is imparted to the free layer magnetization, which balances the energy loss by damping, resulting in steady state precessional motion. For the opposite sign of current the dynamics are damped by both the intrinsic damping and the spin torque, resulting in a quiescent magnetization state. Hence, even though the equilibrium magnetization state is highly symmetric, dynamics are expected for only one sign of $I_{dc}$ (corresponding to electrons flowing from the free layer to the fixed layer), as we experimentally observe. In the modeling we assume a value of the asymmetry parameter of $\Lambda = 1.5$, but the qualitative features reported here are robust with respect to variations of this value, excluding values that negate the existence of the asymmetry, $\Lambda \approx 1$. In general, values of $\Lambda$ that increase the asymmetry in g($\psi$) yield similar magnetization trajectories but at reduced current densities because the spin torque imparts a larger net energy to the free layer for a given current.

The magnetization trajectories and corresponding time traces of the individual components of the magnetization, computed from the single-domain modeling for several values of current, are given in Fig. 2. At the lowest current values the modeling shows that the free layer undergoes roughly circular precession at a constant angular velocity about the axis perpendicular to the film plane (z-axis), resulting in a nominally sinusoidal (single frequency) output. As the current is increased to intermediate values, the amplitude of the oscillation grows while the precession remains quasi-circular about the z-axis and the time-dependence of $M_{free,x}$ (the relevant component for the GMR readout mechanism) remains roughly sinusoidal, as determined by the time traces (see Fig. 2b) and their Fourier transforms (inset). At the higher current values where the precession angle approaches 90º (roughly > 10 mA) the precession remains quasi-circular but the precession axis becomes canted with respect to the z-axis ($\approx$ 10º-15º). This canting correlates with the time-dependence of $M_{free,x}$ strongly deviating from sinusoidal behavior, thereby inducing significant spectral harmonics in the frequency domain (Fig. 2c). This evolution from a single frequency output to an output having strong harmonics at higher current density is in qualitative agreement with the data shown in Fig. 1(a), which shows the emergence of spectral harmonics at the highest current levels. The simulations show that as the current is increased further the frequencies of the dynamics continually decrease and become more non-sinusoidal, and that at the highest currents $M_{free}$ becomes static, pointing in a current-dependent direction in the plane of the film. We are unable to compare this prediction with the data because the devices fail for currents much higher than those shown in Fig. 1.

As discussed above, in the absence of significant harmonic content in the device output, the single-domain simulations of this geometry suggest that the magnetization dynamics are well-described as circular precession about the axis perpendicular to the film plane (*z*-axis) having a constant angular frequency. Neglecting intralayer exchange effects due to the finite extent of the excitation [11], the frequency of precession can then be written as

$$f = \frac{\gamma \mu_0}{2\pi}(H + (H_k - M_s)\cos(\theta)) \qquad (1)$$

where γ is the gyromagnetic ratio and the precession amplitude θ is the angle between $M_{free}$ and the z-axis (Fig. 2a). Since larger current values will increase the angle of precession, red-shifting dynamics with increasing current are expected when $H_k > M_s$, as seen in Fig 1. This is distinctly different than when $H_k$ is smaller than $M_s$ and an applied field is used to induce the free layer to point out of the film plane, as has been investigated previously [7]. In that case, dynamics with increasing amplitude lead to the blue-shifting dynamics with increasing currents (a result that is also consistent with Eq. 1).

In Fig. 3 we plot the onset frequency of the DC induced dynamics as a function of $\mu_0 H$ applied perpendicular to the film plane, along with the ferromagnetic resonance (FMR) frequency of the device excited through the AC spin torque [12,13] for comparison. The linear fit to the spin torque FMR (ST-FMR) data in Fig. 3 gives a value for $\mu_0(H_k - M_s) \approx 0.064$ T through Eq.1. This is in agreement with the value of 0.068 T determined from magnetometry measurements and hard-axis resistance vs. field measurements. As seen in the figure, the precession frequencies from the DC driven dynamics are roughly equivalent to the ST-FMR frequencies, with the two varying by less than 150 MHz for any given field, implying that the frequencies at the onset of DC driven dynamics correspond to small angle precession. The inset in the figure shows an individual ST-FMR scan at an applied field of $\mu_0 H = 0.2$ T, and $I_{dc} = $ 1mA. From similar ST-FMR measurements taken as a function of $I_{dc}$ on this and analogous devices, we determine an $I_{dc} = 0$ mA damping parameter $\alpha = 0.025$-$0.030$, in agreement with previous results [14].

Using Eq. 1 and the value of $\mu_0 (H_k - M_s) = 0.064$ T, we can relate the device oscillation frequency to the angle of precession, which is shown in Fig. 4a, based on the data in Fig. 1b. This analysis indicates that the dynamics begin at small precessional angles (<10º) and grow to larger angles as the current through the device increases. It also suggests that the amplitude of the dynamics appear to exceed 90º at the highest current values. However, this is likely an artifact of the calculation. At the highest current values (largest angles) the appearance of the second harmonic signal in the data of Fig. 1a and the simulations suggest that the dynamics are not properly described as having a constant angular velocity. Hence, in that regime Eq. 1 is no longer valid and will not yield correct values for the angle of precession. Because of this, we consider the reported angles to be most valid at low current values and least so at the highest ones.

The direct translation between the precessional frequency and angle enables the estimation of the variations in the precessional amplitude $\Delta\theta$ from the measured values of $\Delta f$. Theoretical [15] and experimental work [16] suggest that thermally-induced amplitude fluctuations and their ensuing phase noise are the dominant sources of line width broadening in these devices. A rigorous analysis relating the line width to the amplitude fluctuations in the devices is beyond the scope of the present paper. Instead, we consider a simpler model in which we take the device to have an instantaneous frequency determined by the value of $M_z$ (through Eq. 1) and the ensuing line width as a measure of the "wandering" of the precessional amplitude about its average value. In this case, the line width can be related to a nominal value for $\Delta\theta$ through the relation:

$$\Delta f = \frac{\partial f}{\partial I_{dc}} \left( \frac{\partial \theta}{\partial I_{dc}} \right)^{-1} \Delta\theta \ . \qquad (2)$$

Using Eq. 2 and the data shown in Figs 1 and 4a, the derived values of $\Delta\theta$ in this device (for $\mu_0 H$ = 0.25 T) are shown in Fig. 4b. At onset, $\Delta\theta$ is about 5º, decreases to below 2º over most of the current range, and then increases at the highest current values. The increase in $\Delta\theta$ at the largest values of $I_{dc}$ reflects the associated increase in the line width shown in Fig. 1c, but in this regime the reported values should be taken only as estimates. The values of $\Delta\theta$ are only valid over the same range as Eq. 1, since it is used to determine the precessional angles. Nonetheless, the ability to quantitatively connect the precession frequencies and amplitudes through the model enables quantitative estimates of the amplitude fluctuations in these devices. We present this analysis as a point of comparison for future theoretical efforts in relating the line width to amplitude and phase fluctuations in STNO devices.

The device output power also varies with current in accordance with macrospin modeling. Within the limitations discussed above, we interpret the dynamics as corresponding to quasi-circular precession about the z-axis having precession angles that grow with increasing current. In Fig 5, we plot the measured device power output (corrected for cable loss ($\approx$ 2 dB), bias-tee loss, and amplifier gain) as a function of $I_{dc}$. For comparison, we also include the theoretical value for the measured power $P_{Theory}$ for circular precession, which uses the derived relationship between $I_{dc}$ and $\theta$ shown in Fig. 4a. The theoretical values include the power loss associated with the impedance mismatch between the 10 $\Omega$ device and the 50 $\Omega$ measurement circuitry:

$$P_{Theory} = \frac{(I_{DC}\Delta R)^2}{8} \frac{R_{Load}}{(R_{Device} + R_{Load})^2} \quad (3)$$

where $\Delta R$ is the change in the device resistance associated with the particular magnetization trajectory induced by $I_{dc}$, $R_{load} = 50$ $\Omega$ is the input impedance of the amplifier, and $R_{Device}$ is the average effective device impedance at the frequency of interest (in our devices this has been measured to be approximately equal to the DC resistance at these frequencies). However, Eq. 3 does not account for any reflection between our on-chip waveguide structure and the 50 $\Omega$ cabling, which has been experimentally estimated to be roughly 0.5 - 1.0 dB. As seen in the figure, the theoretical and measured values for the output power are in reasonable quantitative agreement. A more rigorous comparison is difficult because we have not precisely calibrated for frequency dependent standing-wave resonances due to the impedance mismatch between the device and the rest of the circuit, which are evidenced by the pronounced dip and peak in the power at roughly 9.5 mA and 11.25 mA, respectively. However, the quantitative agreement of the derived and measured powers supports our interpretation of the induced dynamics discussed above.

In the following paragraphs we examine the dependence of the onset current on the applied field. In Fig. 6, we plot $I_c$, the current at which precessional motion is first detected, as a function of $\mu_0 H$ applied along the *z*-axis up to 0.35 T. A linear fit to the data yields $I_c = (10.7 \pm 0.5)$ (mA/T) $\mu_0 H + (1.18 \pm 0.1)$ mA. We note that for applied fields much larger than $\approx 0.7$ T, $I_c$ no longer increases linearly with $\mu_0 H$ but instead decreases with increasing field. This indicates that the fixed layer magnetization has been pulled out of the film plane enough to cause a significant change in the direction of the spin-current polarization. The reorientation of the CoFe layer results in significantly different dynamical modes than those discussed above, and will be the subject of a different publication. Here, we will limit our analysis to fields below 0.35 T in order to ensure that the in-plane fixed layer approximation used in the analytic theory discussed below is appropriate.

Assuming that the fixed layer lies along the x-axis, an analysis of the Landau-Lifshiftz-Gilbert-Slonczewski equation [11] results in the expression for the critical current as a function of the applied field:

$$I_c = C(H + (H_k - M_s)) + I_{loss}, \qquad (4)$$

where $C = \dfrac{4\alpha e M_s V_{eff}}{P\mu_0} \dfrac{\left(1+\dfrac{1}{\Lambda^2}\right)^2}{\left(1-\dfrac{1}{\Lambda^2}\right)}$, $e$ is the electron charge, and $V_{eff}$ is the nominal volume undergoing precession [17]. The first term in Eq. 4 corresponds to the critical current within the macrospin approximation, and the second term, $I_{loss}$, accounts for additional loss mechanisms associated with the nanocontact geometry. In principle, $I_{loss}$ needs only to account for spin-wave radiation from the contact area [11], but it could also include strongly diffusive current transport/spin accumulation in the actual devices, device-dependent scattering processes, and current shunting through the Cu spacer layer. In the following, we assume that $I_{loss}$ does not depend on $\mu_0 H$, which is appropriate for each of these loss mechanisms. In this case, the individual parameters in Eq. 4 are directly determined through the measurements presented above. Fixing $\mu_0(H_k - M_s)$ at the measured value of 0.064T from the data of Fig 3, we determine $C = 10.7$ mA/T through the slope $dI_c/d\mu_0 H$ and $I_c$ ($\mu_0 H = 0$ T) = 1.18 mA by the intercept of the fit. In principle, the only unknown parameter in $C$ is the value of $\Lambda$, which is a measure of the angular dependence of the spin-transfer torque. Using the measured value of $\alpha = 0.03$, a value of $V_{eff}$ from the free layer thickness and the intended exposure diameter of the contact, an assumed value of $P = 0.35$ (generally consistent with transition metal ferromagnets [18]), and a measured value of $\mu_0 M_s = 0.98$ T (through SQUID magnetometry measurements), we find a value of $\Lambda = 1.5$. In order to quantify the errors associated with this estimate, we determine a range of values for $\Lambda$ by varying the nominal contact diameter between 60 nm and 80 nm and the value of the polarization $0.3 < P < 0.4$, which gives $1.3 < \Lambda < 1.7$. Because of possible difference between the nominal and real contact diameters and the uncertainly in $P$, these values for $\Lambda$ are only first estimates. Nevertheless, based on this methodology, the accuracy of the value for $\Lambda$ can be improved upon and extended to other systems once additional information is available on Co/Ni and other perpendicularly magnetized materials.

From the same data we can also estimate the quantity $I_{loss}$. The analysis below is similar to the one previously carried out for mechanical point contact studies [19], but here we are able to more directly control the size of the fabricated devices and measure the onset of precessional dynamics directly. Using the experimentally determined values of $I_c$ ($\mu_0 H = 0$ T) = 1.18 mA, $dI_c/d(\mu_0 H)$ = 10.7 mA/T, and $\mu_0(H_k - M_s) = 0.064$ T, we find $I_{loss} = (0.5 \pm 0.1)$ mA. We note that this technique of measuring $I_{loss}$ does not depend on assumed values for the device properties (*e.g.,* the device volume, $P$, or $\Lambda$), since the slope and intercept of the $I_c$ vs. $H$ data are directly measured. In the tens of devices measured with similar ($H_k$ - $M_s$), the value of $I_{loss}$ varies typically between about 0.3 mA – 2 mA, which is an important fraction (≈ 30-60 %) of the measured value of $I_c$ ($\mu_0 H$ = 0 T).

We note that our measured room-temperature values for $I_c$ likely differ from the onset current for $T$ = 0 K and $\theta$ = 0°, the conditions under which Eq. 4 is derived. This discrepancy is unavoidable since the measurements are performed at room temperature and a finite amplitude of precession is required for electrical detection. The finite temperature will tend to decrease $I_c$ as compared to its $T$ = 0 K value [20], whereas the requirement that the precession amplitude be sufficient to overcome the noise floor of the instrumentation will tend to inflate the measured values of $I_c$. Hence, there is a possible systematic error introduced when comparing the measured values of $I_c$ with Eq. 4. We do not have a robust method for estimating the decrease of the measured $I_c$ associated with finite temperature effects, but recent experimental work suggests that it is roughly 20% [21]. However, the agreement between the onset frequencies determined through the DC driven measurements and those of the ST-FMR measurements shown in Fig. 3 suggests that the measured values of $I_c$ well approximate the (small angle) onset of dynamics.

In the following we compare our results to the simplest case in which $I_{loss}$ is associated exclusively with spin wave radiation losses in the nanocontact geometry and the measured values of $I_c$ are equivalent to their zero temperature and zero amplitude values, in which case $I_{loss}$ = $(23Aet)/(h\ g(\pi/2))$, where $A$ is the exchange energy density and $t$ is the free layer thickness [11]. An experimentally determined value of $A$ for Co/Ni multilayers is not available in the literature, so we use the nominal values of $A = 1\text{-}2\ (10^{-11})$ J/m, representing a typical range of values for CoNi alloys [22]. Thus, we find that the expected $I_{loss}$ ranges between 3 mA and 10 mA, depending on the value used for $A$ and the particular functional form considered for $g(\psi)$ [10,11]. This range of theoretical values for $I_{loss}$ is generally larger than those determined experimentally and could indicate that the spin wave radiation loss in our devices is not isotropic within the film plane, as is assumed in Ref. [11]. For instance, the radiation could be suppressed due to reflections from defects in the film or localization due to a combination of the red-shifting behavior of the dynamics and the experimental requirement of a finite amplitude of precession for their detection [23]. Furthermore, the radiation could have an angular dependence due to the Oersted fields associated with $I_{dc}$ [24, 9]. In order to be more quantitative, an experimentally determined value for $A$ in the Co/Ni multilayer system is needed, as are theoretical models which include the effects of finite temperature on the determination of $I_c$.

In summary, we have measured spin-torque-driven dynamics in STNO devices having a perpendicularly magnetized free layer and in-plane magnetized fixed layer in fields < 0.35 T applied perpendicularly to the film plane. The frequencies of the precession and the measured output powers are both in accordance with the single-domain model of this geometry and indicate that their dynamics increase from small angle precession at onset to large angle (≈90º) quasi-circular, but canted, precession at larger currents. By comparing the measurements with the single-domain model we are able to determine the precession angle as a function of frequency and estimate the thermally induced variations of the precession amplitude in these devices. Additionally, from measurements of the onset current as a function of applied field we are able to estimate the value for the spin-torque asymmetry parameter for this system, as well as directly determine the excess loss associated with the nanocontact device geometry. This system offers the ability to generate large angle dynamics, directly accessing the details of the angular dependence of the spin torque parameters, and measuring the losses associated with spin wave radiation in STNO devices. A more detailed understanding of the dynamics, particularly at the highest currents, will require further development of strongly non-linear and strongly out of equilibrium theories. With the use of materials having larger out-of-plane anisotropies, this geometry may lead to the development of current tunable, large angle, high frequency nanoscale oscillators that do not need an externally applied field to operate.

We thank A. Wickenden, C. Fazi, and B. Huebschman for  measurements of the device impedance at the frequencies discussed here and M. Kuperferling, P. Kabos, M. T. Wallis, R. Heindl, T. Cecil, M. Schneider, A. Kos, and T. Silva for helpful discussions and assistance in the measurements. This work was partially supported by the NIST Office for Microelectronic Programs, the NIST Innovation in Measurement Science program, and the Marie Curie Fellowship Association.

Fig. 1 (a) (color online) Two-dimensional plot showing the device oscillation frequency and output power as a function of $I_{dc}$ for an external field of 0.25 T applied perpendicular to the film plane. The power spectral density (PSD) is shown in a linear color scale. The second harmonic signal (see text) does persist up to the highest current values, but at significantly reduced power density, as shown in the inset. The magnitude has been corrected for amplifier gain and cable loss. All spectral data were taken on a time scale of roughly 1 s. (b) Oscillation frequency and (c) and line width (FWHM) of the data shown in part (a) as determined by Lorentzian fits to the spectra.

Fig. 2 (a) Magnetization trajectories from the single domain simulations for several current values. (b,c) Simulated time traces showing each component of the magnetization for $I_{dc} = 5.5$ mA and 18 mA, respectively. The effects of the trajectories canting away from the z-axis can be seen through the time variation of $M_z$ and the Fourier transforms of $M_x$ (insets), the relevant component for the GMR readout mechanism. For currents less than roughly 10 mA, the time traces show nominally sinusoidal variations of $M_x$ (as is evidenced by the $I_{dc} = 5.5$ mA example, where all higher harmonics are less than 3% of the fundamental). As the current is increased the time dependence significantly deviates from sinusoidal, as seen in (c) for $I_{dc} = 18$ mA, where the second harmonic is roughly 20% of the fundamental. The simulations are for a device having dimensions of 100 nm x 100 nm x 3 nm, an external field of 0.25 T applied along the $z$-axis, a damping parameter $\alpha = 0.03$, and a simulated temperature of 5 K.

Fig. 3 (a) Onset frequencies of the DC driven dynamics (circles) along with the resonance frequencies (squares) determined through ST-FMR measurements as a function of applied field. (Inset) Plot showing an example of a frequency swept ST-FMR scan taken in an external field of 0.2 T and $I_{dc} = 1$ mA.

Fig .4 (a) Mapping of the measured device oscillation frequency to the derived oscillation amplitude $\theta$. The error bars are determined through standard error propagation techniques and the uncertainty in the experimentally determined value of the $\mu_0 H = 0$ T intercepts of the fits to the DC driven and ST-FMR data in Fig. 3. The increased error bars at small angles reflect the sensitivity of the error propagation associated with the arc-cosine function near 0°. (b) Calculated values of amplitude fluctuations $\Delta\theta$ as a function of precessional amplitude $\theta$. The values are calculated through Eq. 2 in which $\Delta f$ is taken as the FWHM values reported in Fig. 1 (c).

Fig. 5. Plots of the measured and theoretical output power as functions of $I_{dc}$. The theoretical plot is determined through Eq. 3 and the data shown in Fig. 4 (a), as described in the text. The maximum $\Delta R$ for this device is 100 mΩ.

Fig. 6. The measured onset current $I_c$ as a function of the external field applied perpendicular to the film plane along with a linear fit to the data. The error bars determined by the current interval between adjacent spectral measurements are ± 0.05 mA and are smaller than the data points in the plot.

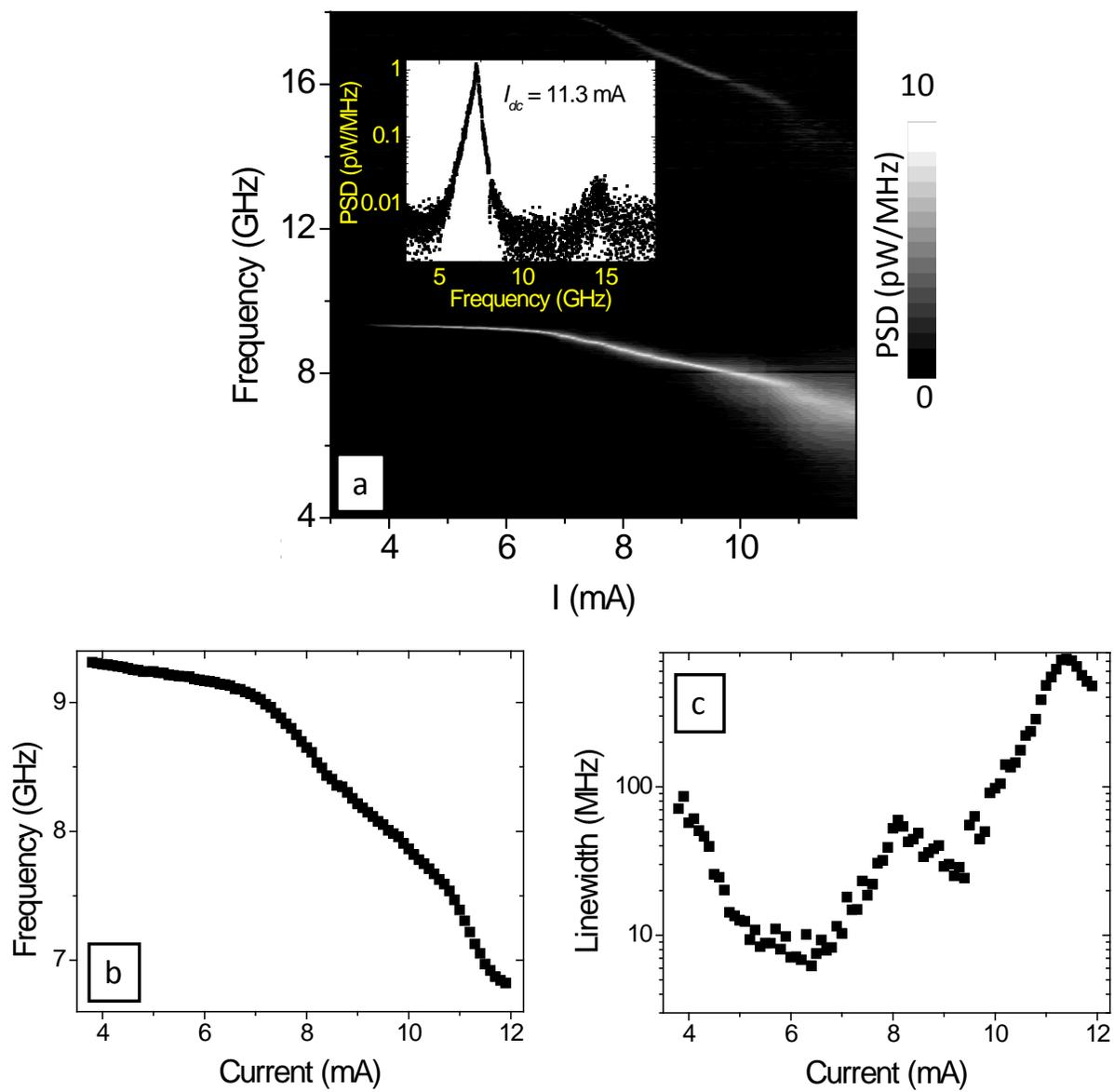

Figure 1

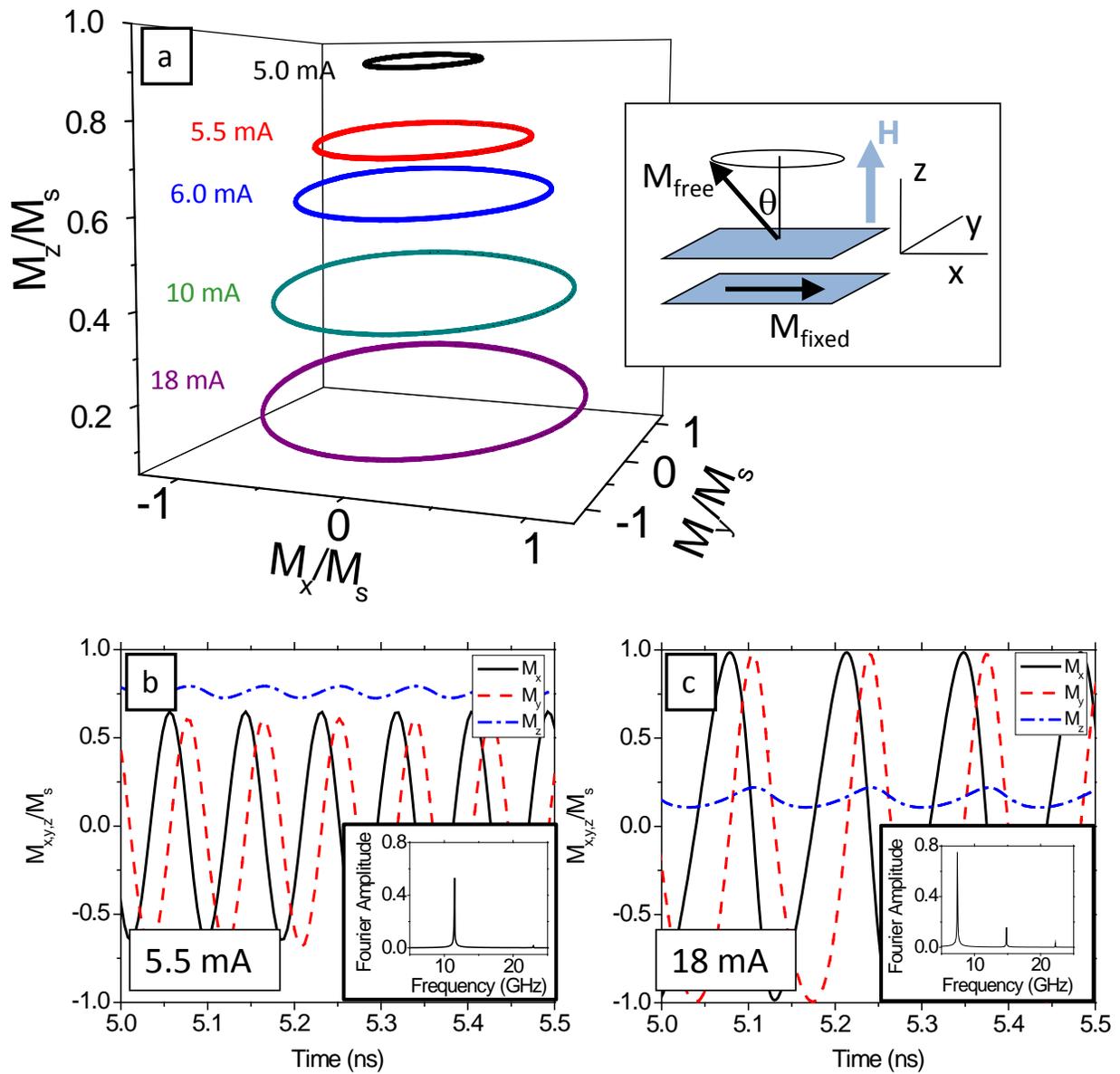

Figure 2

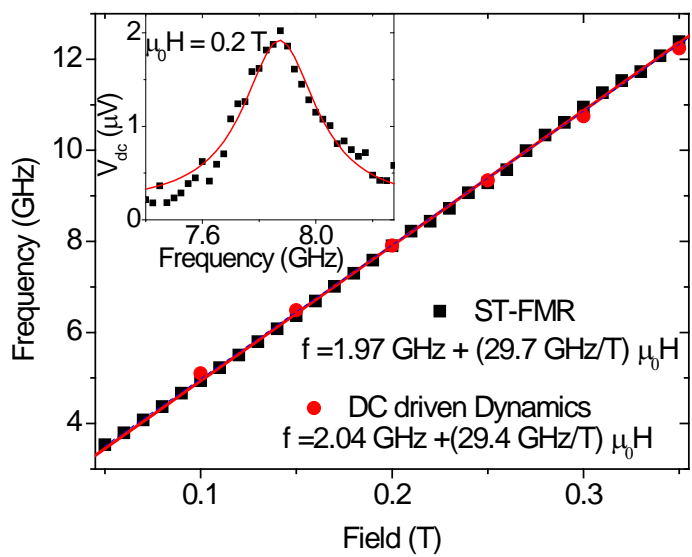

Figure 3

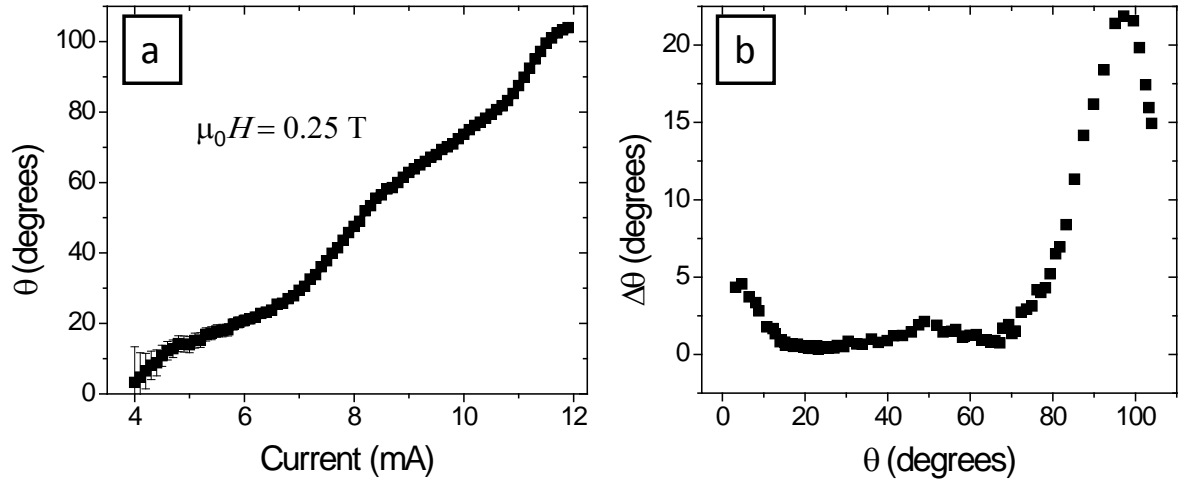

Figure 4

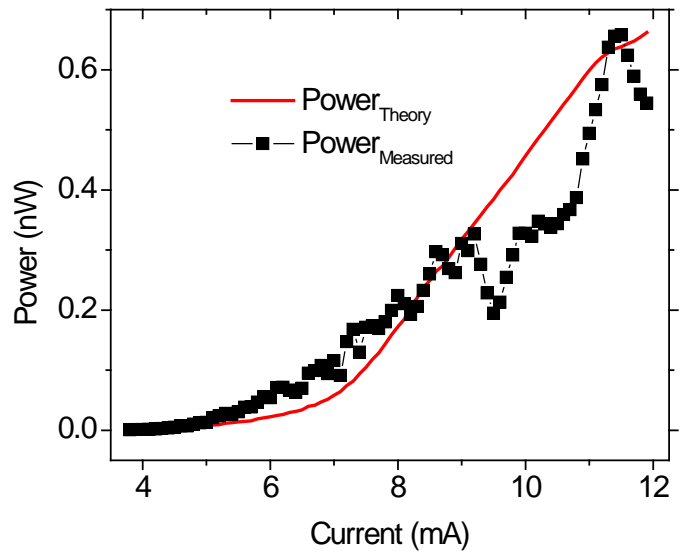

Figure 5

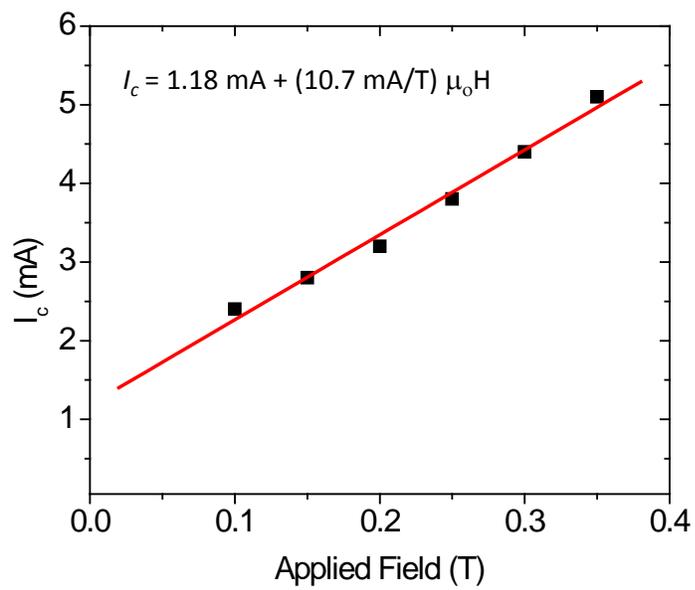

Figure 6